\documentclass[12pt,aps,prd,amsmath,amssymb,showpacs,floats,floatfix,nofootinbib]{revtex4-1}
\usepackage[colorlinks=true, linkcolor=red, citecolor=blue,CJKbookmarks=true]{hyperref} 
\usepackage{amsmath,amsfonts,amssymb}
\usepackage{graphicx}
\usepackage{color}
\usepackage[dvipsnames,svgnames,x11names]{xcolor}
\usepackage{slashed} 
\usepackage{url}
\usepackage{subfigure}
\usepackage{threeparttable}
\usepackage{multirow} 
\usepackage{setspace} 
\graphicspath{{figs/}}  
\def\TeV{\mathrm{TeV}} 
\def\GeV{\mathrm{GeV}} 
\def\kpc{\mathrm{kpc}} 
\def\Mpc{\mathrm{Mpc}} 

\makeatother
\allowdisplaybreaks 

\begin{document}

\setstretch{1.2} 
\title{Prospect for dark matter annihilation signatures from gamma-ray observation of dwarf galaxies by LHAASO}

\author{Dong-Ze He$^1$}
\author{Xiao-Jun Bi$^{2,3}$}
\author{Su-Jie Lin$^2$}\email{linsj@ihep.ac.cn}
\author{Peng-Fei Yin$^2$}
\author{Xin Zhang$^{1,4}$}\email{zhangxin@mail.neu.edu.cn}
\date{\today}
\affiliation{$^1$Department of Physics, College of Sciences, Northeastern University, Shenyang 110819, China}
\affiliation{$^2$Key Laboratory of Particle Astrophysics, Institute of High Energy Physics, Chinese Academy of Sciences, Beijing 100049, China}
\affiliation{$^3$School of Physical Sciences, University of Chinese Academy of Sciences, Beijing 100049, China}
\affiliation{$^4$Center for High Energy Physics, Peking University, Beijing 100080, China}

\begin{abstract}
The Large High Altitude Air Shower Observatory (LHAASO) is a next-generation observatory for high energy gamma rays and cosmic rays with wide field of view. It will detect gamma rays with high sensitivity in the energy range from 300 GeV to 1 PeV. Therefore, it is promising for LHAASO to search for the high-energy gamma rays induced by dark matter (DM) self-annihilation in dwarf spheroidal satellite galaxies (dSphs), which are ideal objects for the DM indirect detection. In this work, we investigate the LHAASO sensitivity to DM self-annihilation signatures for 19 dSphs and take the uncertainties on the spatial DM distribution of dSphs into account. We perform a joint likelihood analysis for the 19 dSphs and find that the LHAASO sensitivity to the DM annihilation cross section will reach $\mathcal{O}(10^{-24})\sim \mathcal{O}(10^{-25})$ cm$^3$ s$^{-1}$ at the mass scale above TeV for several annihilation modes, which is larger than the canonical thermal relic cross section by a factor of 10 to 100.
\end{abstract}

\maketitle

\section{Introduction}\label{sec:intro}
A lot of compelling astrophysical and cosmological observations have indicated the existence of nonbaryonic cold dark matter,
which constitutes nearly 25\% of the energy budget of the Universe \cite{Adam:2015rua}.
DM is essential in the evolution of the Universe and the formation of large-scale structures. However, in spite of the acknowledged existence of DM, we still have a poor understanding about its fundamental properties as an elementary particle, and do not know its interaction with the Standard Model particles other than gravity. To reveal these mysteries, many new physics models have been proposed in the literature, among which a popular kind of the DM candidate is the weakly interacting massive particles (WIMPs) \cite{Jungman:1995df,Bergstrom:2000pn,Bertone:2004pz}.

WIMPs could either annihilate\footnote{In this paper, we only focus on Majorana WIMPs instead of the Dirac WIMPs, thus all the ``annihilation'' mentioned in the following would refer in particular to self-annihilation.} or decay in astrophysical systems today, and then produce steady and energetic Standard Model particles, such as protons/antiprotons, electrons/positrons, neutrinos, and photons. One kind of current DM identification method, namely the indirect DM detection, is just to search for such nongravitational signals
and to further reveal the physical properties of DM particles. Particularly, the observations of high-energy gamma-ray emissions produced by DM annihilation, either monoenergetic (from direct annihilation) or with a continuum of energies (through the cascade decays or final-state radiations), are of great interest and importance, because the process of gamma-ray propagation is not deflected by the interstellar magnetic field and can naturally trace back to the sites where DM annihilations occur. As the DM annihilation rate is proportional to the square of DM density distributions, these gamma-ray signatures would be preferentially generated in the DM dominated regions, and then can be detected by terrestrial and satellite experiments.

Hitherto, quite a number of works have been performed to study the gamma-ray signatures from DM annihilation in many different dark-matter-rich astrophysical objects, such as galaxy clusters \cite{Ackermann:2010rg}, the galactic halo \cite{Hooper:2011ti,Ackermann:2012rg,Abazajian:2012pn,Abdo:2010nc,Ackermann:2012qk,Weniger:2012tx,Ackermann:2013uma}, galactic DM substructures \cite{Zechlin:2011kk,Ackermann:2012nb,Zechlin:2012by}, etc. Since there is no significant gamma-ray excess to date that has been confirmed, the stringent upper limits on the DM annihilation cross section have been reported in the literature \cite{Abdo:2010ex,Ackermann:2011wa,GeringerSameth:2011iw,Cholis:2012am,GeringerSameth:2012sr,Mazziotta:2012ux,Baushev:2012ke,Huang:2012yf,Ackermann:2013yva,Ackermann:2015zua,Tsai:2012cs}.
Among these astrophysical objects, the dwarf spheroidal satellites (dSphs), known as large galactic substructures surrounding the Milky Way, are considered to be the most promising and ideal laboratories for the indirect DM detection. First, the mass-to-light ratios in dSphs can be of very large order of magnitude, which suggests that they are significantly DM-dominated systems. Second, dSphs are also expected to be relatively free of gamma-ray emission from other astrophysical sources, since they have little or no recent star formation activity and detected ionized gas \cite{Mateo:1998wg,Grcevich:2009gt}. These outstanding advantages could extremely simplify the interpretation of a gamma-ray excess potentially detected in the direction of a dSph.

During the past several decades, the achievements in gamma-ray astronomy either in the GeV range with space-borne instruments or in the TeV region with ground-based detectors, have produced extraordinary advances in high-energy astrophysics. However, for the gamma-ray sky in the energy range above a few tens of TeV, the past and present telescopes can only record few photons, which makes this energy region almost completely unexplored.
Under this circumstance, strong interest is addressed to the development of next-generation instruments, which are able to make more precise observations in a more extended energy range with a high sensitivity.
Currently, the most sensitive detectors for very high energy (VHE) gamma rays are imaging air Cherenkov telescopes (IACTs), such as H.E.S.S \cite{Hinton:2008zz}, VERITAS \cite{Furniss:2012zz}, and MAGIC \cite{Albert:2007xh}. But the sensitivity of IACTs would be limited by their small field of view (FOV) and short operation duty cycle. The ground-based air shower particle detectors, such as Tibet-AS$\gamma$ and ARGO-YBJ may overcome those disadvantages of IACTs, but the poor background rejection power still limits their sensitivities. One of the reasonable methods to improve the sensitivity of the ground-based array detectors is to detect muons in the shower, such as the muon detector of the Tibet-AS$\gamma$ experiment.

Most importantly, the under-construction Large High Altitude Air Shower Observatory (LHAASO) project \cite{Cao:2010zz,Cao:2014rla} will become a continuously operated gamma-ray telescope at energies from $\sim$ 300 GeV to 1 PeV and open a new window for the gamma-ray detection.
LHAASO is designed to maintain a high sensitivity as well as a strong background rejection power ($\sim1\%$) and a large FOV ($\sim$~2 sr) simultaneously.
Therefore, through the VHE gamma-ray observation from dSphs by LHAASO, it is very promising to detect the DM annihilation signatures or set strong limits on the properties of heavy DM.
From such a point of view, we investigate the prospects for detecting the DM annihilation signature by the LHAASO observations of 19 dSphs. We also take the uncertainties of the $J$-factor of dSphs into account \cite{Geringer-Sameth:2014yza,Hutten:2016jko} and study the impact of these uncertainties on the LHAASO sensitivity. To derive a reasonable sensitivity, the simulated data of LHAASO considering the background rejection power are also used.

It is difficult to detect the annihilation signals from thermally produced DM particles by LHAASO or other cherenkov detectors, as the annihilation cross section required by the thermal relic density is much smaller than the experimental reach. However, if DM particles are produced nonthermally, they can still have a large annihilation cross section. There have been many nonthermal mechanisms proposed in the literature. For instance, the DM particles can be produced as the decay products of heavy particles, Q balls, or cosmic strings (see, e.g., Refs.~\cite{Moroi:1999zb,Fujii:2002kr,Jeannerot:1999yn} and references therein). It is meaningful to search for the signatures from these DM particles in the future indirect detection experiments.

This paper is organized as follows. In Sec. \ref{sec:LHAASO}, we give a brief introduction to the LHAASO experiment. In Sec.   \ref{sec:method}, we discuss the calculation of gamma-ray flux from DM annihilation and introduce the methods of simulating the gamma-ray observation and of calculating the limits on DM annihilation cross section. We show the LHAASO sensitivities to the DM annihilation cross section and make comparison with other experimental results in Sec. \ref{sec:dm-limit}. Finally, the conclusion is given in Sec.  \ref{sec:conclu}.

\section{LHAASO Observatory}\label{sec:LHAASO}
LHAASO is a hybrid cosmic-ray and gamma-ray observatory located at 4410 m above sea level near Daocheng, Sichuan province, China ($100^{\circ}.01$E, $29^{\circ}.35$N). The LHAASO experiment is composed of a square kilometer particle detector array (KM2A), a water Cherenkov detector array (WCDA), a wide field Cherenkov telescope array (WFCTA), and a high threshold shower core detector array (SCDA).

KM2A is primarily designed for the detection of VHE gamma rays ($E\gtrsim10$ TeV). The surface array consists of 5195 scintillator electron detectors with 1 $\rm m^{2}$ each and a spacing of 15 m. The large effective area of approximately kilometers squared of the surface detectors could provide enough exposure for the photons with high energies. With the purpose of rejecting the cosmic-ray background, 1171 underground muon detectors with 36 $\rm m^{2}$ each and a spacing of 30 m will be built under the surface detector array, with the total active area being up to 40,000 $\rm m^{2}$. For the energies above 50 TeV, KM2A will achieve a background-free detection of photons and make LHAASO the most sensitive observatory around the world.

WCDA, located at the center of the KM2A array, is attributed to the gamma-ray detection in the energy range $\lesssim$ 20 TeV. It is composed of four water pools with 150 $\times$ 150 $\rm m^{2}$ each, and the total active area is 90,000 $\rm m^{2}$, which is 4.5 times larger than that of the High-Altitude Water Cherenkov (HAWC) experiment. In addition, WFCTA and SCDA are dedicated to measure the cosmic-ray spectra of individual composition, providing a multiparameter measurement in order to better distinguish between different compositions. Specifically, WFCTA can detect the longitude evolution of a cosmic-ray shower, and SCDA can detect the shower components near the core.

\begin{figure*}
\includegraphics[width=0.70\textwidth]{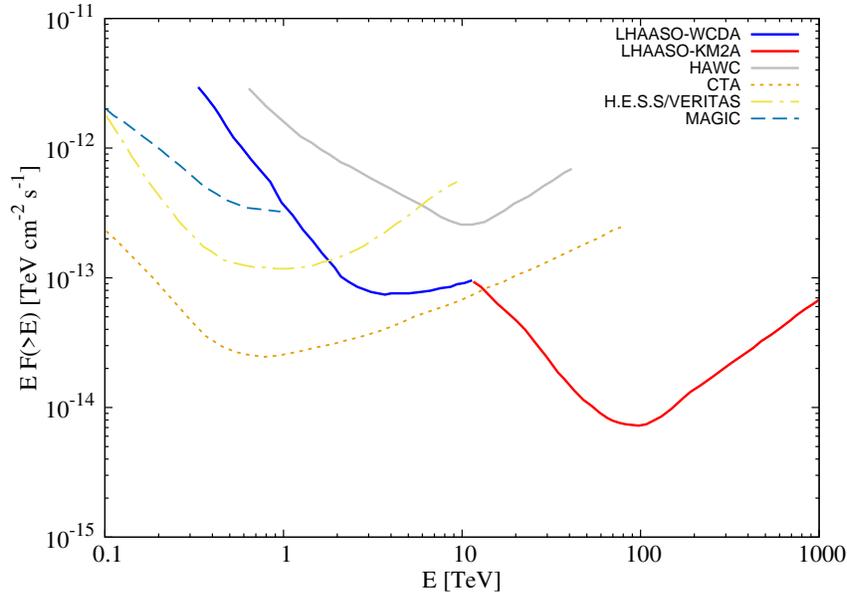}
\caption{Simulated integral sensitivity of LHAASO for a Crab-like source, compared with the sensitivities of other experiments \cite{Acharya:2013sxa,Cao:2014rla}. The observation times are 1 year and 50 hours for wide field of view detectors and IACTs, respectively.}
\label{fig:integral-sensi}
\end{figure*}

The gamma-ray sensitivity of LHAASO to a Crab-like source is shown in Fig. \ref{fig:integral-sensi}~\cite{Cao:2014rla}. In this figure, the exposure time is one year for air shower array experiments and 50 hours for IACTs. It is shown that for energies above 20 TeV LHAASO will be the most sensitive gamma-ray experiment in the world. The three major goals of LHAASO are 1) surveying the VHE gamma-ray sky with a sensitivity of $\sim1\%$ of the Crab Nebula flux, 2) precisely measuring the cosmic-ray spectrum of individual composition at the knee region and beyond, and 3) exploring the new physics frontiers. For the layout of the detectors and a more detailed description of the experiment, we refer the reader to Refs. \cite{Cao:2010zz,Cao:2014rla}.

The most relevant detectors for gamma-ray detection of LHAASO are KM2A and WCDA. Thanks to the large area of the array KM2A and the high capability of background rejection, LHAASO can reach sensitivities for gamma rays with energies above $\sim$ 30 TeV, about 100 times higher than that of current experiments, offering the possibility to monitor the gamma-ray sky up to 100 TeV for the first time,
and thus is preferably effective for the detection of gamma rays from Galactic source.
The threshold energy of WCDA can be as low as $\sim$ 300 GeV and thus it could be effective for the extragalactic sources.
As it is still under construction, in this paper we exclusively focus on the discussion about the prospects of searching for heavy annihilating DM particles above TeV from the LHAASO gamma-ray observation of dSphs.

\section{Analysis method}\label{sec:method}

\subsection{Gamma-ray fluxes from DM annihilation in dSphs}\label{sec:gamma-flux}
The expected gamma-ray flux from DM annihilation is calculated with the utilization of not only the astrophysical properties of the potential DM distribution, but also the properties of the initial- and final-state particles in different annihilation channels. For self-conjugate DM particles, the gamma-ray integral flux from the pair annihilation of DM particles in a dSph (point-like source) can be given by
\begin{equation}\label{func:anni-flux}
\Phi=\frac{1}{4\pi}\frac{\langle\sigma v\rangle}{2m_{\chi}^{2}}\int^{E_{\rm max}}_{E_{\rm min}}\frac{dN_{\gamma}}{dE_{\gamma}}dE_{\gamma} \times J,
\end{equation}
where $m_{\chi}$ is the DM mass, $\langle\sigma v\rangle$ is the thermal average velocity-weighted DM annihilation cross section, $\frac{dN_{\gamma}}{dE_{\gamma}}$ is the differential spectrum of prompt photons resulting from DM annihilation, and the integration is for each energy bin between $E_{\rm min}$ and $E_{\rm max}$. Note that $\frac{dN_{\gamma}}{dE_{\gamma}}$ should be a sum of all the photons from any possible DM annihilation final states according to the DM model. However, in this analysis, we only hypothesize the gamma-ray contribution from a certain annihilation channel through the use of the {\tt PPPC4DM} package \cite{Cirelli:2010xx,Ciafaloni:2010ti}.

In Eq.~(\ref{func:anni-flux}), $J$ represents the astrophysical ``$J$-factor",  which is the integral of the DM density squared along the line of sight ($\rm l.o.s$) distance $x$ in the region of interest (ROI)
\begin{equation}\label{func:jfactor}
J=\int_{\rm source}d\Omega\int_{\rm l.o.s} dx\rho^{2}(r(\theta,x)),
\end{equation}
where $\Omega$ denotes the solid angle of the observation region over which the $J$-factor is calculated,
and can be expressed as $\Delta \Omega=2\pi\times[1-\cos\alpha_{\rm int}]$, where $\alpha_{\rm int}$ is the integration angle.
The density profile $\rho{(r)}$ describes how the density of an astrophysical system varies with the distance $r$ from its center, which is given by
\begin{equation}
  r(\theta,x)=\sqrt{R^{2}-2xR\cos\theta+x^{2}},
\end{equation}
where $R$ is the distance from the Earth to the source center and $\theta$ is the angle between the orientation to the center of source and the line of sight.

The DM density profile of dSphs can be derived from the kinematic observation of stellar velocities through the use of the Jeans equation (see e.g. Refs. \cite{Evans:2003sc,Strigari:2007at,Martinez:2009jh}). There are several systematic and statistical uncertainties in the determination of the DM density profile and $J$-factor. For instance, the stellar surface brightness and velocity anisotropy profiles are required in the Jeans analysis, but these profiles have not been uniquely determined. For the ultra-faint dSphs with large $J$-factors that are of interest to the indirect detection, the lack of kinematic data would also induce large statistical uncertainties in the $J$-factor. The discussions on the systematic uncertainties in the $J$-factor of dSphs can be found in Refs. \cite{Lefranc:2016dgx} and the references therein.

In this work, we take the calculated mean values of $J$-factor and their statistical uncertainties of 19 dSphs in Table \ref{Table:dsphs}~\cite{Geringer-Sameth:2014yza,Hutten:2016jko}.
In Ref.~\cite{Geringer-Sameth:2014yza}, the $J$-factors are calculated by using the Zhao profile \cite{Hernquist:1990be,Zhao:1995cp}, which is parameterized as\
\begin{equation}
\rho(r)=\frac{\rho_s}{(r/r_s)^\gamma[1+(r/r_s)^\alpha]^{(\beta-\gamma)/\alpha}},
\end{equation}
where $\rho_s$, $r_s$, $\alpha$, $\beta$, and $\gamma$ are free parameters. For $(\alpha, \beta, \gamma)=(1,3,1)$, this profile becomes the standard Navarro-Frenk-White (NFW) profile. However, this profile can also describe the cored halo for $\gamma\sim 0$. In principle, the free parameters in the profile can be extracted in the Jeans analysis of dSph kinematics and then be used to derive the median value and credible interval of the $J$-factor. The other typical profile used in Ref.~\cite{Hutten:2016jko} is the Einasto profile~\cite{Graham:2005xx}, which is parameterized as
\begin{equation}
\rho(r)=\rho_{-2}\left\{ -\frac{2}{\alpha}\left[\left(\frac{r}{r_{-2}}\right)^\alpha-1\right] \right\},
\end{equation}
where $\rho_{-2}$, $r_s$, $r_{-2}$, and $\alpha$ are free parameters. As discussed in Ref.~\cite{Bonnivard:2014kza}, the results given by these two typical kinds of profiles are in very good agreement.

Notice that there are two sets of $J$-factors in Refs.~\cite{Geringer-Sameth:2014yza,Hutten:2016jko}, one of which is derived with a constant integration angle $\alpha_{\rm int}=0.5^\circ$ and the other of which is derived with the integration angle which equals the maximum angular radius $\alpha_{\rm int}=\theta_{\rm max}$. The maximum angular radius $\theta_{\rm max}$  of each source is indicated by its outermost member star in the current observation. As the annihilation contribution of the gamma-ray signals which is proportional to the DM squared density ($\rho^2$) would concentrate more around the center of the dSphs, a smaller integration angle would always lead to a larger signal-to-background ratio in this work.
Therefore, in Table.~\ref{Table:dsphs}, we choose the $J$-factor within a smaller integration angle as $\alpha_{\rm int}=\min\{\theta_{\rm max},0.5\}$.

In this work, as we mainly focus on the gamma ray below $E_{\rm up}\equiv20\,\TeV$, the absorption effect of the signals, resulting from the interactions with the cosmic microwave background (CMB) and the Galactic photon field, is supposed to be negligible. For more detailed discussions, see Appendix~\ref{gamma_absorption}.

\begin{table*}
\caption{\label{Table:dsphs}The astrophysical properties of 19 selected dSphs within the LHAASO FOV.
The listed columns for each dSph are the name, right ascension (RA.), declination (DEC.),
effective time ratio ($r_{\rm eff}$) showing the fraction of the observation time during which the corresponding zenith angle is smaller than $60^\circ$,
maximum angular radius ($\rm \theta_{max}$) associated with the dSph's outermost member star,
and $J$-factor.
The $J$-factor and $\rm \theta_{max}$ of the dSphs are taken from Ref. \cite{Geringer-Sameth:2014yza}. However, for the four dSphs marked with asterisks of which the $J$-factors are not given in Ref. \cite{Geringer-Sameth:2014yza}, we utilize the calculated results from Ref. \cite{Hutten:2016jko}.}
\small\centering
\begin{tabular}{ccccccccccccccc} \hline \hline
              &&
          RA. &&
          DEC.&&
          $r_{\rm eff}$&&
          $\theta_{\rm max}$&&
          $\log_{10}J_{\rm obs}$&&\\
		
       Source&&
		(deg)&&
		(deg)&&
		     &&
		(deg)&&
		($\GeV^{2}\rm cm^{-5}$)\\\hline

		$\rm Bo\ddot{o}tes ~I$&&
		$210.02$&&
		$14.50$&&
		$0.352$&&
		$0.47$&&	
		$18.2\pm0.4$\\

		Canes Venatici $\rm I$&&
		$202.02$&&
		$33.56$&&
		$0.398$&&
		$0.53$&&
		$17.4\pm0.3$\\
		
		Canes Venatici~$\rm II$&&
		$194.29$&&
		$34.32$&&
		$0.399$&&
		$0.13$&&
		$17.6\pm0.4$\\
		
		Coma~Berenices&&
		$186.74$&&
		$23.90$&&
		$0.377$&&
		$0.31$&&
		$19.0\pm0.4$\\
		
		Draco&&
		$260.05$&&
		$57.92$&&
		$0.442$&&
		$1.30$&&
		$18.8\pm0.1$\\

		Draco II$^{\star}$&&
		$238.20$&&
		$64.56$&&
		$0.451$&&
		$-$&&
		$18.1\pm2.8$\\
		
        Hercules&&
		$247.76$&&
		$12.79$&&
		$0.348$&&
		$0.28$&&
		$16.9\pm0.7$\\

        Leo I&&
		$152.12$&&
		$12.30$&&
		$0.346$&&
		$0.45$&&
		$17.8\pm0.2$\\

        Leo II&&
		$168.37$&&
		$22.15$&&
		$0.372$&&
		$0.23$&&
		$18.0\pm0.2$\\

        Leo IV&&
		$173.23$&&
		$-0.54$&&
		$0.303$&&
		$0.16$&&
		$16.3\pm1.4$\\

        Leo V&&
		$172.79$&&
		$2.22$&&
		$0.314$&&
		$0.07$&&
		$16.4\pm0.9$\\

        Pisces II$^{\star}$&&
		$344.63$&&
		$5.95$&&
		$0.327$&&
		$-$&&
		$16.9\pm1.6$\\

        Segue 1&&
		$151.77$&&
		$16.08$&&
		$0.357$&&
		$0.35$&&
		$19.4\pm0.3$\\

        Sextans&&
		$153.26$&&
		$-1.61$&&
		$0.299$&&
		$1.70$&&
		$17.5\pm0.2$\\

        Triangulum II$^{\star}$&&
		$33.32$&&
		$36.18$&&
		$0.403$&&
		$-$&&
		$20.9\pm1.3$\\

        Ursa Major I&&
		$158.71$&&
		$51.92$&&
		$0.432$&&
		$0.43$&&
		$17.9\pm0.5$\\

        Ursa Major II&&
		$132.87$&&
		$63.13$&&
		$0.449$&&
		$0.53$&&
		$19.4\pm0.4$\\

        Ursa Minor&&
		$227.28$&&
		$67.23$&&
		$0.455$&&
		$1.37$&&
		$18.9\pm0.2$\\

        Willman 1$^{\star}$&&
		$162.34$&&
		$51.05$&&
		$0.430$&&
		$-$&&
		$19.5\pm0.9$\\
    \hline\hline
	\end{tabular}
    \end{table*}
    
\subsection{Events at LHAASO}\label{sec:events}

The ground-based experiment LHAASO is impinged by secondary particles from the Extensive Air Shower (EAS) induced by cosmic rays.
By monitoring the generated Cherenkov light in the water Cherenkov detectors (WCDs), one can estimate the cosmic particles' energy and determine the orientations from which the initial cosmic particles arrived.
Different kinds of initial cosmic particles (hadron/gamma) would lead to different energy distributions in the WCDs across the array.
For example, a gamma-ray shower results in a smoother energy distribution, whereas a hadron shower leads to a clumped distribution across the WCDs.
Using this feature, we can discriminate the gamma signals from the backgrounds induced by the hadronic primary incoming cosmic-ray particles.

Since LHAASO is still under construction, at present we could only make a mimic observation.
The procedure is as follows. First, we calculate the expected counts of background events resulting from the incoming cosmic-ray particles. Then, we assume that there are no significant signals from DM annihilation and make a Gaussian sampling around the background event counts $B$ to get a mimic total observational counts $N$.

The energy resolution of WCDA is about tens of percents.
We use wide energy bins ($E_{\rm max}/E_{\rm min}=3$ for each bin 
in the energy range $700\,\GeV\sim20\,\TeV$
) in order to suppress the systematic errors induced by the reconstructed energy dispersion.
In each energy bin, the background events $B$ from the hadronic cosmic-ray particles can be calculated by
\begin{equation}\label{func:bkg}
  B=\int^{E_{\rm max}}_{E_{\rm min}}\int_{\Delta\Omega}\int_0^T\zeta_{cr}\cdot\Phi_{p}(E)\cdot A_{\rm eff}^{p}(E,\theta_{\rm zen}(t))\cdot\varepsilon_{p}(E)dtd\Omega dE,
\end{equation}
where $\Phi_{p}(E)$ is the flux of protons in the primary cosmic rays described by a single power-law spectrum, which is best fitted by the observational datasets of experiments ATIC \cite{Panov:2011ak}, CREAM \cite{Yoon:2011aa}, and RUNJOB \cite{Derbina:2005ta}.
As the abundance of the rest particles in the cosmic ray is about 10\% of the proton's abundance, we introduce an additional factor $\zeta_{cr}=1.1$ to take the contributions of other particles into account. The observational time $T$ is taken to be one year.
The integration is calculated in all the energy bins within a cone which is defined as $\Delta\Omega=2\pi\times[1-\cos(\max\{ \alpha_{\rm int}, \theta_c \})]$, where $\theta_c$ is the energy-dependent angular resolution of LHAASO. In fact, $\theta_c$ varies from $2^\circ$ to $0.1^\circ$ with the increased photon energy as described in Fig. 45 of the LHAASO Science White Paper \cite{Bai:2019khm}.

The effective area $A_{\rm eff}^{p}$ is derived from an interpolation calculation in the Science White Paper of LHAASO, which is a function of energy and zenith angle $\theta_{\rm zen}$ \cite{Bai:2019khm}.
DSphs at different declinations correspond to different $\theta_{\rm zen}(t)$ functions, and finally result in different visibilities. To briefly show the visibility of each dSph, we list in Table~\ref{Table:dsphs} the effective time ratio $r_{\rm eff}$, which is determined by the proportion of observation time during which the zenith angle $\theta_{\rm zen}$ is smaller than $60^\circ$.

With regard to the survival ratio $\varepsilon$ in the $\gamma/p$ discrimination on WCDA,
in Ref.~\cite{Zha:2017vcs} the authors have provided an estimation for the quality factor $Q\equiv\varepsilon_{\gamma}/\sqrt{\varepsilon_{p}}$ at various energies, in which the proton survival rate $\varepsilon_{p}$ varies from 0.04\% to 0.11\% when the gamma survival rate $\varepsilon_{\gamma}$ is around $50\%$ for the energies above $0.6\,\TeV$. While as a more conservative choice in this work, we just set $\varepsilon_{p}$ to be $\sim$ 0.278\% when we keep $\varepsilon_{\gamma} \sim40.13\%$.
\footnote{Note that in this work we mainly focus on the gamma photons with energies higher than 700 GeV.
    This energy range corresponds to $N_{\rm fit} > 20$ as shown in Ref. \cite{Zha:2017vcs}; thus, most of the events with $N_{\rm fit} < 20$ are actually ignored here.
    Therefore, although the effective areas we used in this work are obtained with $N_{\rm fit} > 10$ in Ref.~\cite{Bai:2019khm} while the gamma/hadron separation are obtained with $N_{\rm fit}>20$~\cite{Zha:2017vcs}, their inconsistency in our interesting energy region is negligible.
}

The cosmic-ray electron/positron would share the same survival ratio with the gamma ray, which is 40.13\%.
However, their flux above hundreds of $\GeV$ is less than $r_{e/p}\sim0.002$ times the cosmic-ray hadron~\cite{TheDAMPE:2017dtc,Aguilar:2015ooa}. Therefore, for the energy range of LHAASO, the background contribution from electron/positron would always be less than $\varepsilon_{\gamma}r_{e/p}/\varepsilon_p\sim30\%$. Note that the $30\%$ is just an extreme value at the energy $300\,\GeV$, for higher energies this number would rapidly decrease. Thus, we neglect this part of background contribution.

Similarly, the signal events $S$ can be expressed as
\begin{equation}\label{func:sig}
  S=\epsilon_{\Delta\Omega}\int^{E_{\rm max}}_{E_{\rm min}}\int_0^T\Phi_{\gamma}(E)\cdot A_{\rm eff}^{\gamma}(E,\theta_{\rm zen}(t))\cdot\varepsilon_{\gamma}(E)dtdE,
\end{equation}
where $\epsilon_{\Delta\Omega}=0.68$ is the fraction of observed event counts within the angular resolution of the instrument.
Here, we assume that all the dSphs are point-like sources, as the energy-dependent angular resolution of LHAASO is always larger than the typical angular radius scale of the inner region of the dSph, in which DM annihilations have the most important contribution to the gamma-ray flux.
The effective area $A_{\rm eff}^{\gamma}$ is derived from the same procedure as that in Eq. (\ref{func:bkg}), and $\Phi_{\gamma}(E)$ is the flux of gamma photons from DM annihilation as described in Eq. (\ref{func:anni-flux}).

\subsection{Statistic analysis}\label{sec:analysis}
To quantify the gamma-ray excess in a particular sky region, we perform a likelihood ratio test, which is determined by the ratio of the likelihoods under two hypotheses. The test statistic (TS) is calculated by
\begin{equation}\label{func:ts}
  \rm TS=-2\ln\bigg(\frac{\mathcal{L}_{0,{\rm max}}}{\mathcal{L}_{\rm max}}\bigg),
\end{equation}
where $\mathcal{L}_{0,{\rm max}}$ is the maximal likelihood under the null hypothesis without DM contribution, and $\mathcal{L}_{\rm max}$ is the maximal likelihood under the alternative hypothesis with DM contribution, evaluated at the value of the cross section which maximizes the likelihood. The factor of 2 in the definition is for the purpose of causing the distribution of TS values to asymptotically approach a $\chi^{2}$ distribution. Both likelihoods are taken to be Poisson distribution
\begin{equation}\label{func:likelihood-l}
  \mathcal{L}(\textbf{S}|\textbf{B},\textbf{N})=\prod_{i}\frac{(B_{i}+S_{i})^{N_{i}} {\rm{exp}} [-(B_{i}+S_{i})]}{N_{i}!},
\end{equation}
where $i$ denotes the $i$-th energy bin, $S_{i}$ is the sum of expected number of signal counts corresponding to a DM annihilation cross section, $B_{i}$ is the number of expected background counts, and $N_{i}$ is the number of total observed counts.
Since the value of $S_{i}$ is physically restricted to be positive, for the sources within the under-fluctuations of the background (i.e., $S_{i}<0$), the value of $S_{i}$ maximizing the likelihood is expected to be zero, being consistent with no gamma-rays from DM sources. In this case, we can get $\mathcal{L}_{0}=\mathcal{L}_{\rm max}$, leading to a TS value of zero for the under-fluctuations.

We also consider the statistical uncertainty in $J$-factor determination as a nuisance parameter in the likelihood formulation, following the approach in Refs. \cite{Ackermann:2015zua,Fermi-LAT:2016uux}.
The likelihood in all energy bins for the $j$-th dSph can be written as
\begin{equation}
\mathcal{L}_{j}=\prod_{i}\mathcal{L}_{ij}(S_{ij}|B_{ij},N_{ij})\times\mathcal{J}(J_{j}|J_{{\rm obs},j},\sigma_{j}),
\end{equation}
where $i$ and $j$ represent the $i$-th energy bin and $j$-th dSph, respectively. The $J$-factor likelihood term for the $j$-th dSph is assumed to be a Gaussian term
\begin{equation}
\mathcal{J}(J_{j}|J_{{\rm obs},j},\sigma_{j})=\dfrac{1}{{\rm ln}(10)J_{{\rm obs},j}\sqrt{2\pi}\sigma_{j}}\times e^{-[\rm log_{10}(J_{\it j})-log_{10}(J_{obs,{\it j}})]^{2}/2\sigma_{\it j}^{2}},
\end{equation}
where $\rm log_{10}(J_{obs,{\it j}})$ and $\sigma_{j}$ are the observed mean values and corresponding standard deviations. In the practical calculation, the $\rm log_{10}(J_{\it j})$ is chosen to maximize the $\mathcal{L}_{j}$ for given $\langle\sigma v\rangle$ and $m_{\chi}$.

For a $\chi^{2}$-distributed TS, in order to set a one-side $95\%$ confidence level limit, we expect to derive the decreasing likelihood with an increasing number of photons emitted from a potential DM source.
We optimize $\rm \Delta TS=TS-TS_{95}$ = 2.71 corresponding to an alternative hypothesis excluded at $95\%$ C.L. \cite{Rolke:2004mj}
\begin{equation}\label{func:95-limit}
  -2\ln\bigg(\frac{\mathcal{L}_{0,{\rm max}}}{\mathcal{L}_{\rm max}}\bigg)+2\ln\bigg(\frac{\mathcal{L}_{0,{\rm max}}}{\mathcal{L}_{95}}\bigg)=2\bigg(\rm ln\mathcal{L}_{max}-ln\mathcal{L}_{95}\bigg)=2.71.
\end{equation}
Then, we can set 95\% C.L. upper limit on the DM signature flux by requiring that the corresponding log likelihood has decreased by 2.71/2 from its maximum.
After deriving the allowed amount of signal counts $S_{95}$ at $95\%$ C.L., we impose Eqs. (\ref{func:anni-flux}) and (\ref{func:sig}) to derive the corresponding values of $\left\langle \sigma v\right\rangle_{95} $.

For the joint likelihood analysis of many dSphs, the analysis procedure is similar to the single dSph analysis. The combined likelihood of all dSphs becomes
\begin{equation}
\mathcal{L}^{\rm tot}=\prod_{j}\mathcal{L}_{j}.
\end{equation}
By adjusting the number of $\rm \left\langle \sigma v\right\rangle$, we can get $\rm \left\langle \sigma v\right\rangle_{95}$, satisfying $2\bigg(\rm ln\mathcal{L}_{max}-ln\mathcal{L}_{95}\bigg)=2.71$.

\section{LHAASO Sensitivities}\label{sec:dm-limit}
In this section, we describe the LHAASO sensitivity to the DM annihilation cross section through the gamma-ray observation towards dSphs.
The simulated integral flux sensitivity curve of LHAASO project to a Crab-like source is shown in Fig.~\ref{fig:integral-sensi}; the sensitive curves for other projects are also shown in the same figure for comparison \cite{Cao:2014rla}.
We can clearly see that LHAASO is more sensitive at high-energy range above $\sim 10$ TeV than other ground-based projects. This implies that LHAASO will have a better capability to explore the property of heavy DM particles.

\begin{figure*}[!htbp]
{\includegraphics[width=0.45\textwidth]{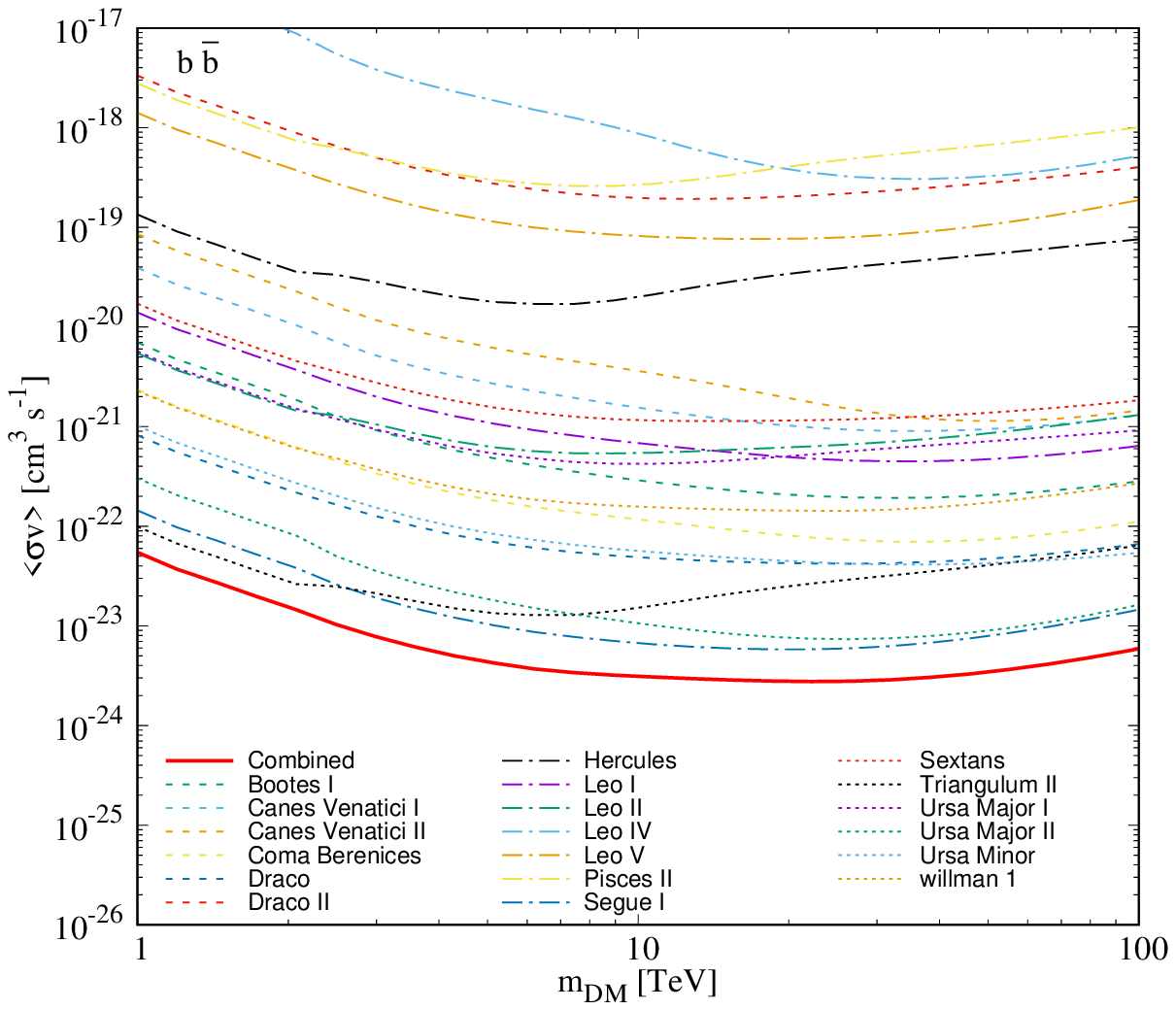}}
{\includegraphics[width=0.45\textwidth]{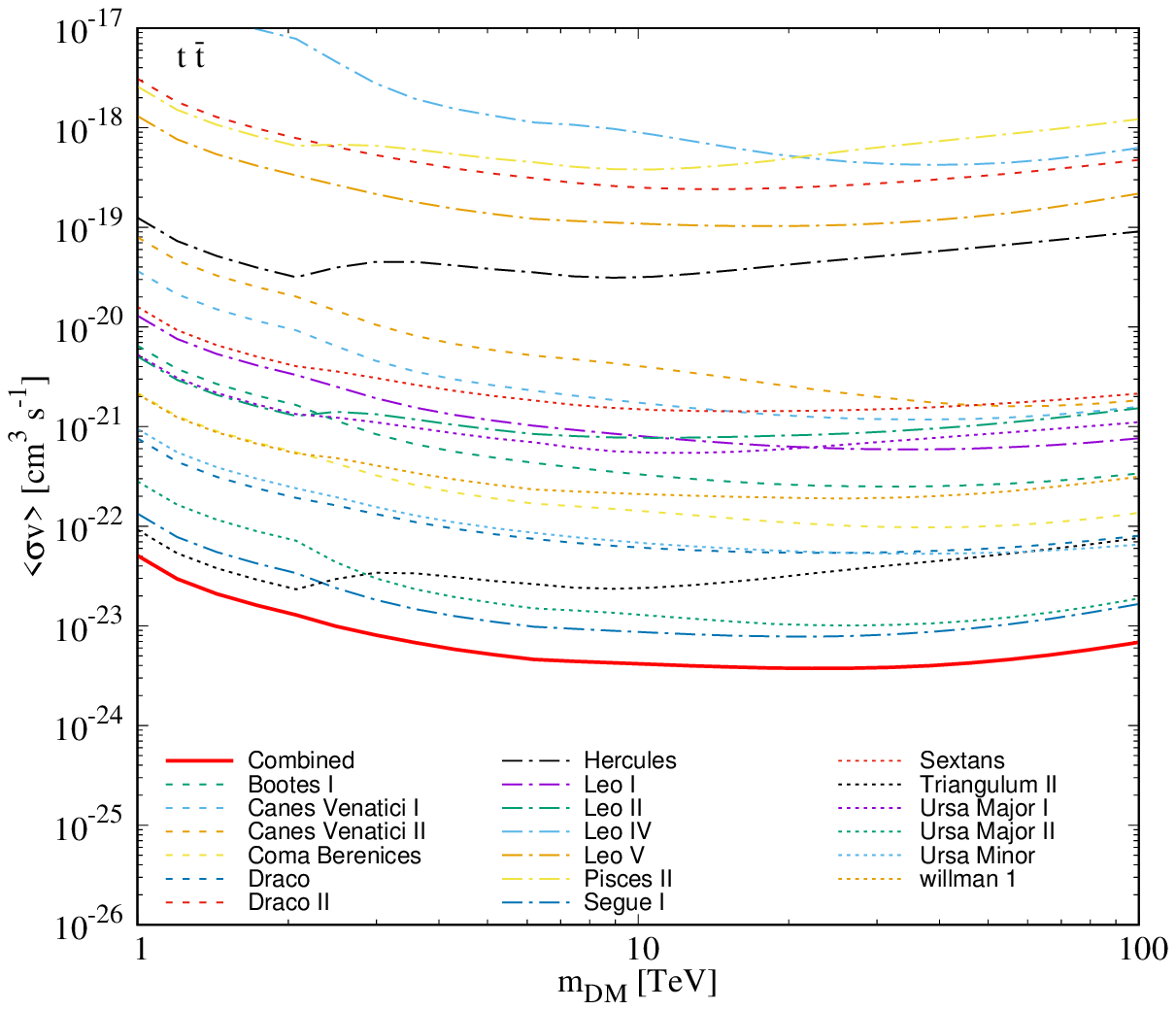}}
{\includegraphics[width=0.45\textwidth]{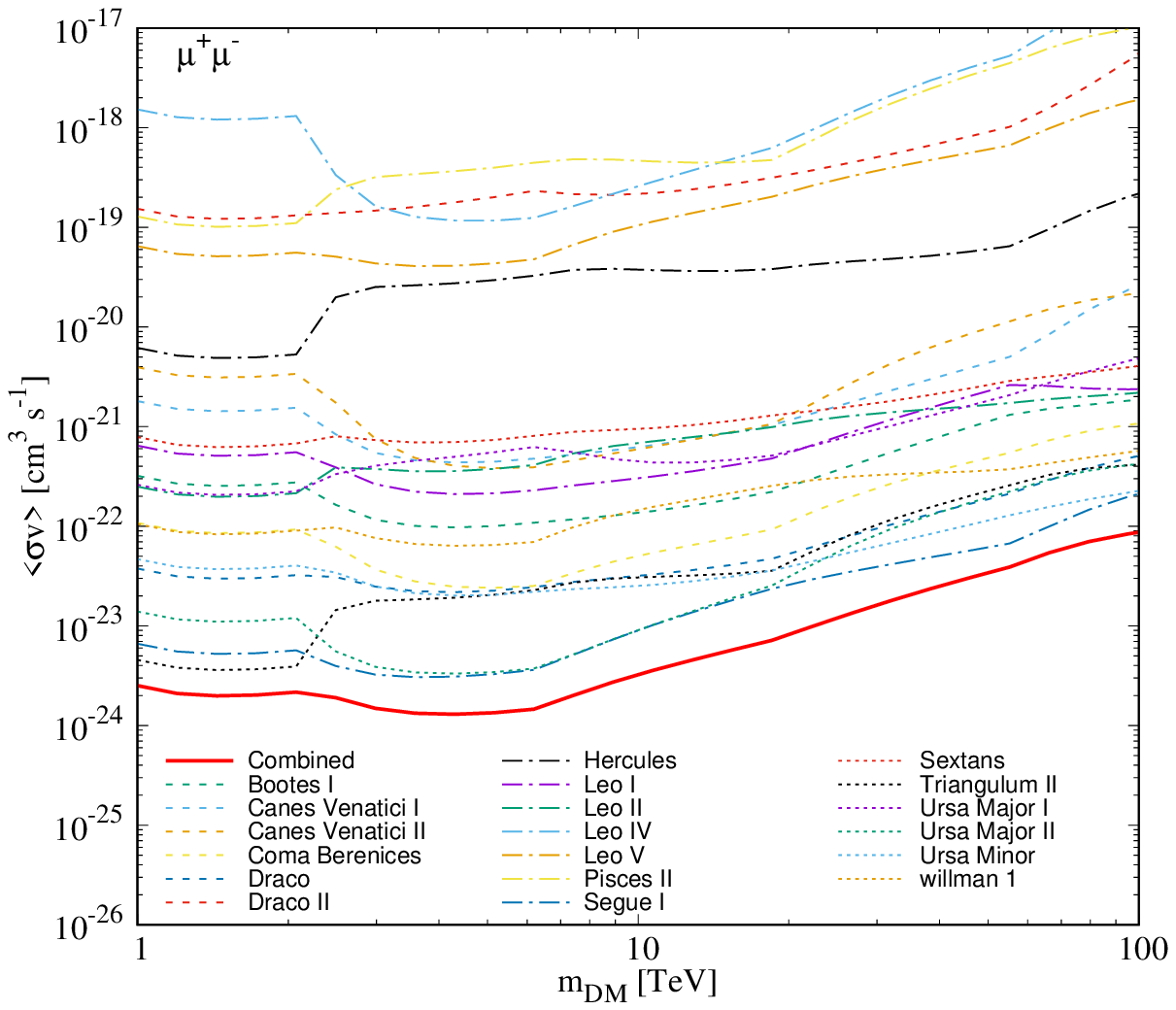}}
{\includegraphics[width=0.45\textwidth]{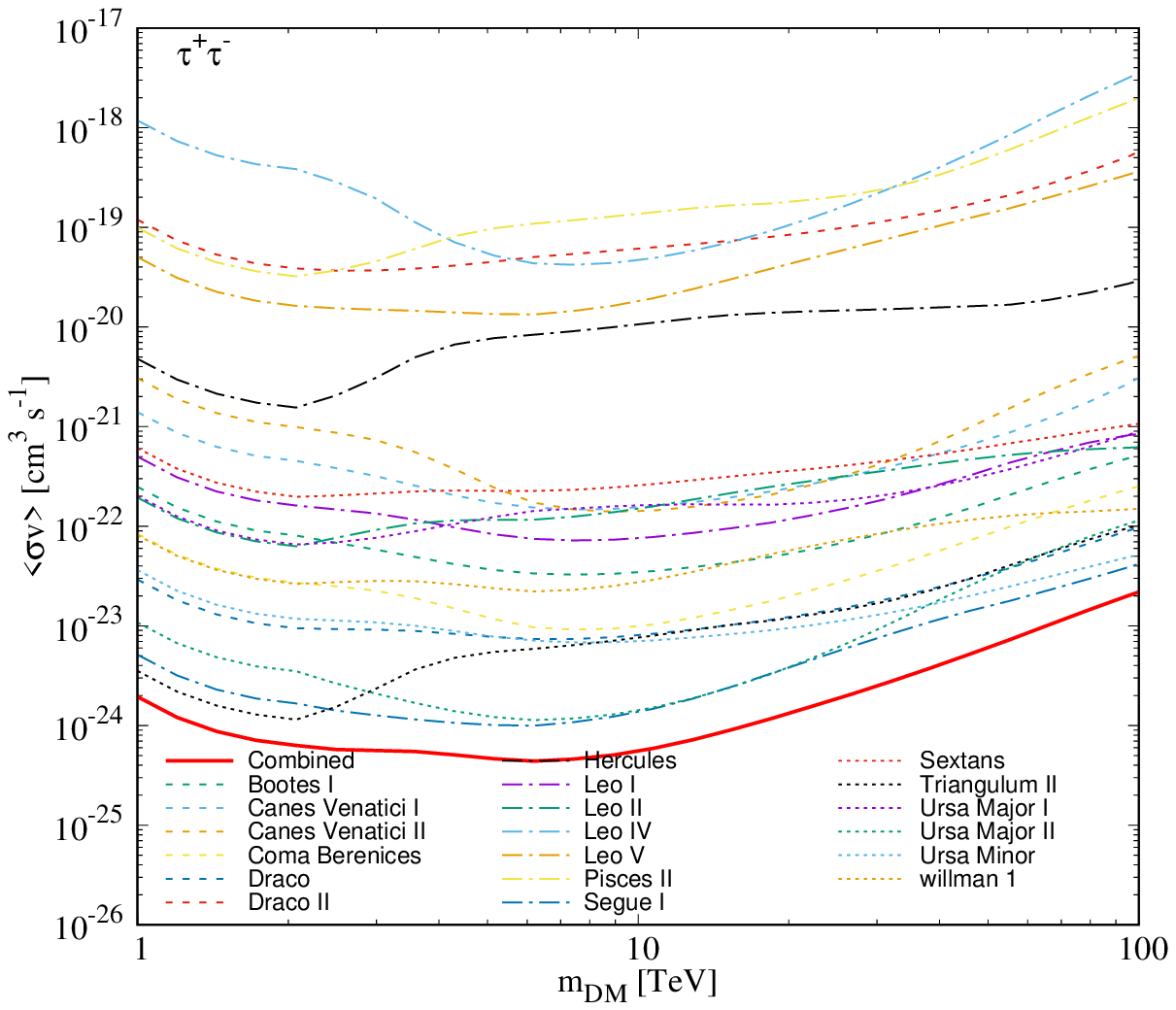}}
{\includegraphics[width=0.45\textwidth]{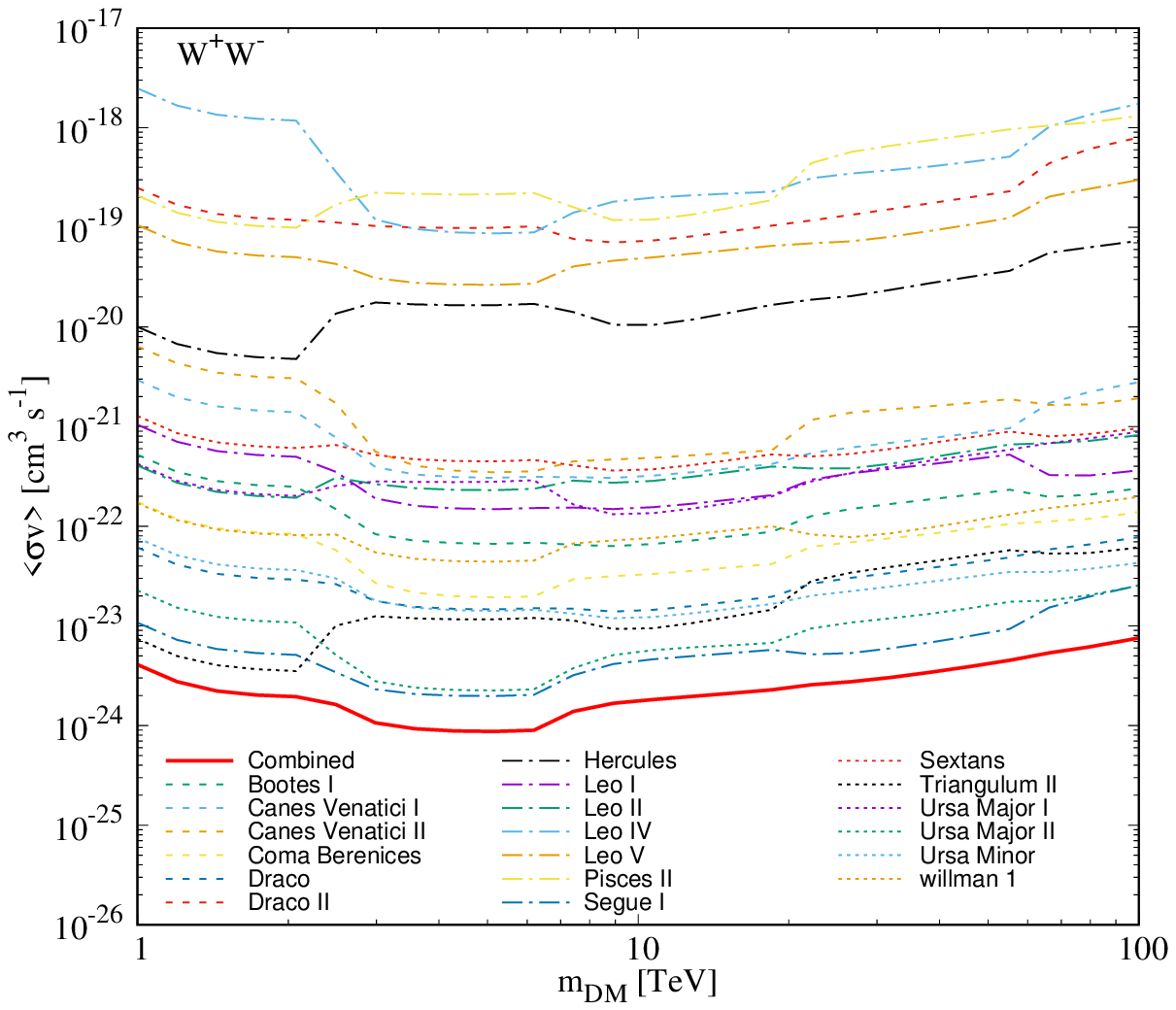}}
\caption{The projected sensitivities to the DM annihilation cross section $\langle\sigma v\rangle$ at $95\%$ confidence level for 19 dSphs within the LHAASO FOV of one year for the $\rm b\bar{b}$, $\rm t\bar{t}$, $\mu^{+}\mu^{-}$, $\tau^+\tau^-$, and $\rm W^+W^-$ annihilation channels. The solid red line represents the combined sensitivity resulting from a joint likelihood analysis, considering the observations of all dSphs.}
\label{fig:Lhaaso-separate}
\end{figure*}

We select 19 dSphs inside the FOV of LHAASO, mean values and uncertainties of the $J$-factor of which are listed in Table \ref{Table:dsphs}. These dSphs are chosen for their favored declination angle for LHAASO and have well-studied dark matter contents. Because of the large FOV of LHAASO (defined in the declination range $-11^{\circ}<\delta<69^{\circ}$), we involved four more dSphs (Draco II, Leo V, Pisces II, and Willman 1) in the analysis, compared with the observation of HAWC \cite{Albert:2017vtb}. In light of the simulated gamma-ray observation of LHAASO, we calculate the sensitivities to the DM annihilation cross section for five annihilation channels $b\bar{b}$, $t\bar{t}$, $\mu^{+}\mu^{-}$, $\tau^+\tau^-$, and $\rm W^+W^-$, as shown in Fig. \ref{fig:Lhaaso-separate}.
The individual sensitivities for each dSph are considered. In addition, the combined sensitivities resulting from a joint likelihood analysis for all the selected dSphs are also exhibited. For the sake of improving the research comprehensiveness, we take the statistical uncertainty of the $J$-factors into account, which would more or less loosen the sensitivity to the gamma-ray signature from dSphs. In spite of this issue, our result is still better than the current upper limit set by HAWC \cite{Albert:2017vtb} by a factor of $2\sim 5$.

There are two reasons for this improvement.
First, the area of WCDA is about 4.5 times larger than that of HAWC, which would improve the sensitivity by a factor of $\sim2.1$.
Second, we have adopted a more efficient $\gamma/p$ discrimination compared with the analysis of HAWC.
In the studies of HAWC~\cite{Capistran:2015dua,Abeysekara:2017mjj}, when $\varepsilon_\gamma$ is set to be $\sim50\%$, $\varepsilon_p$ is always larger than $1\%$ around $1\,\TeV$.
However, the corresponding value for WCDA is expected to be much smaller \cite{Zha:2017vcs}.
Actually, we have adopted a more conservative $\varepsilon_p\sim0.278\%$ compared with Ref. \cite{Zha:2017vcs}. In spite of this, the good $\gamma/p$ discrimination of WCDA still leads to an improvement with a factor at least of $\sim2$.

We find that the combined sensitivity is dominated by the influence of three dSphs with large $J$-factors and favorable locations in the FOV of LHAASO, including Segue 1, Ursa Major II, and Triangulum II. Since LHAASO is located at the latitude of $\sim29^{\circ}$, it would be more sensitive than HAWC to the high-latitude bright sources such as Ursa Major II. Although Triangulum II with almost the largest $J$-factor among all the selected dSphs is very close to the center of the LHAASO FOV, it is not utterly dominant over the combined sensitivity. This is because the statistical uncertainty of the $J$-factor of Triangulum II is large due to the lack of kinematic observational data. Including the uncertainty of the $J$-factor in the joint likelihood analysis would alleviate the overestimation of the combined sensitivity to a great extent. The remaining 16 dSphs do not significantly impact on the combined sensitivity. Despite that some of them are relatively close to the center of the LHAASO FOV, those dSphs have so small $J$-factors that LHAASO is not sensitive enough to them.

\begin{figure*}
	{\includegraphics[width=0.45\textwidth]{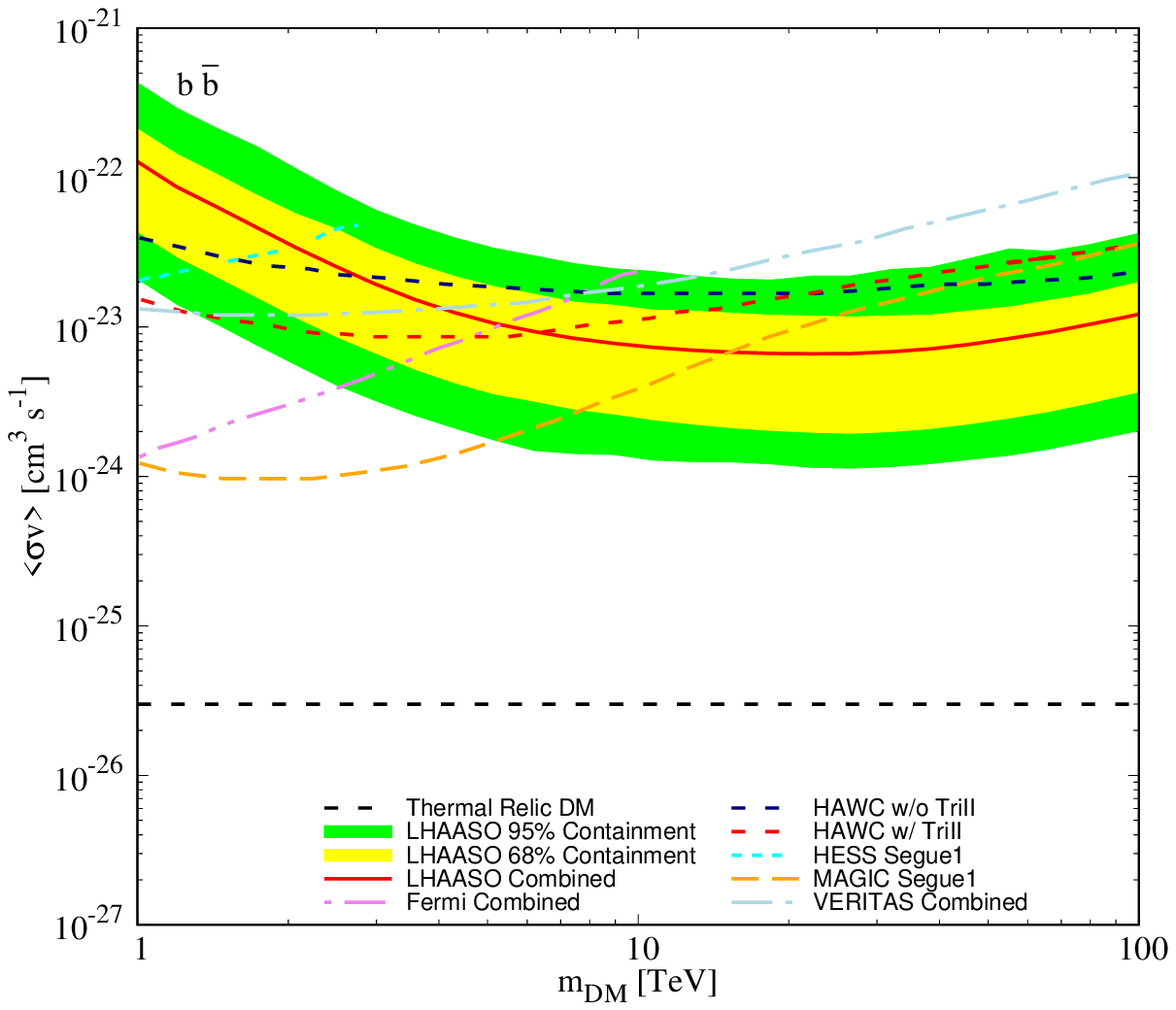}}
	{\includegraphics[width=0.45\textwidth]{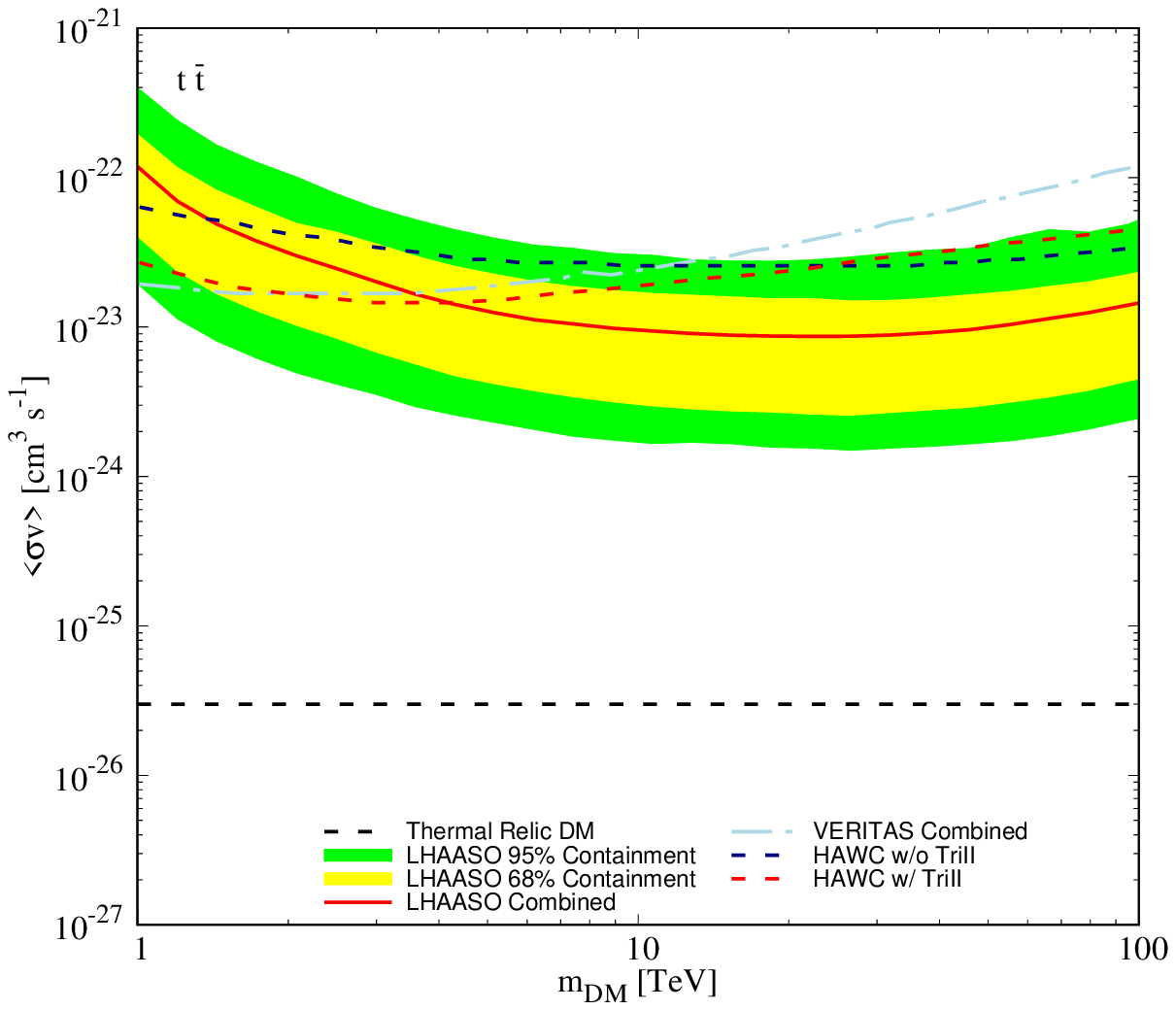}}
	{\includegraphics[width=0.45\textwidth]{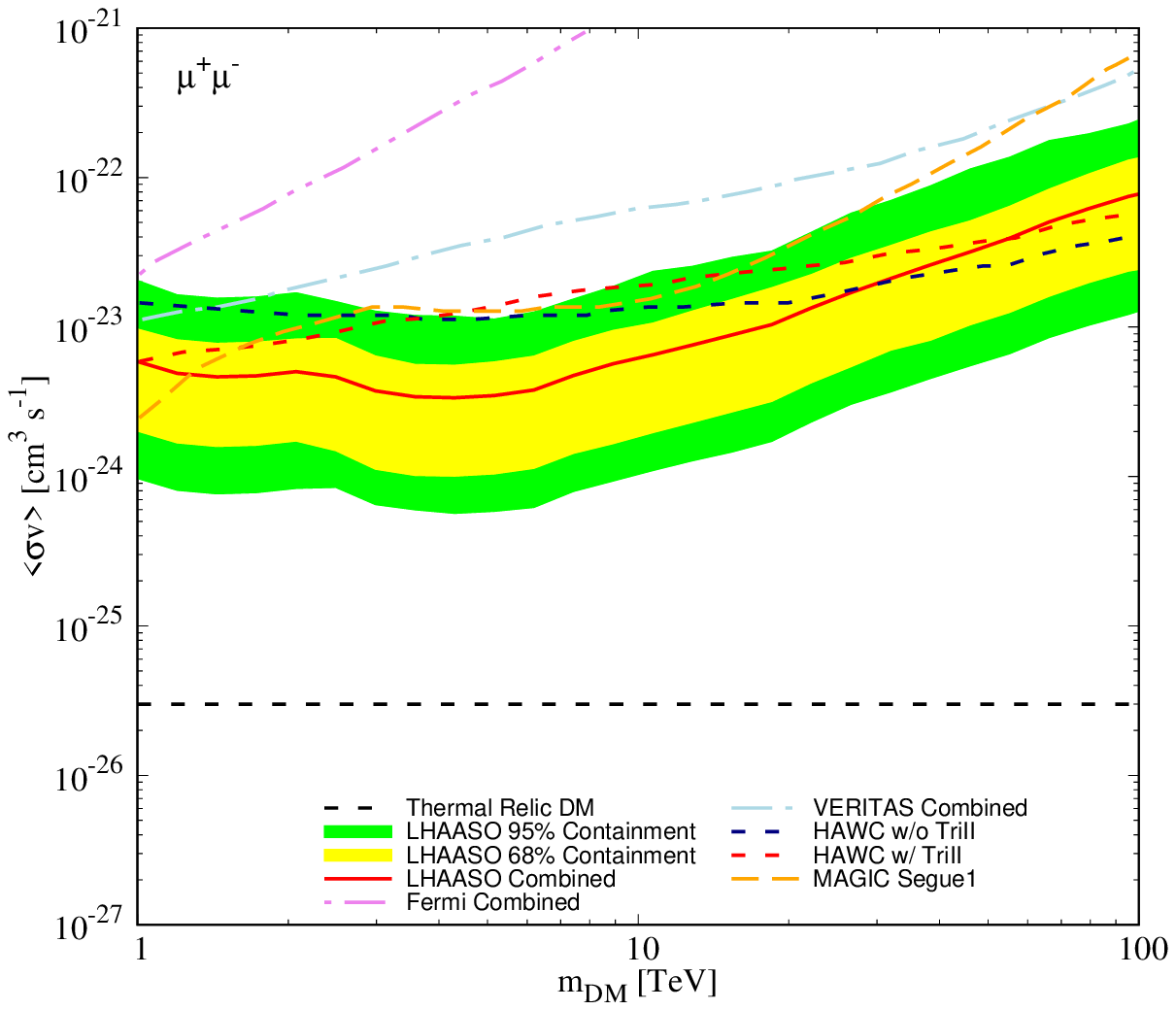}}
	{\includegraphics[width=0.45\textwidth]{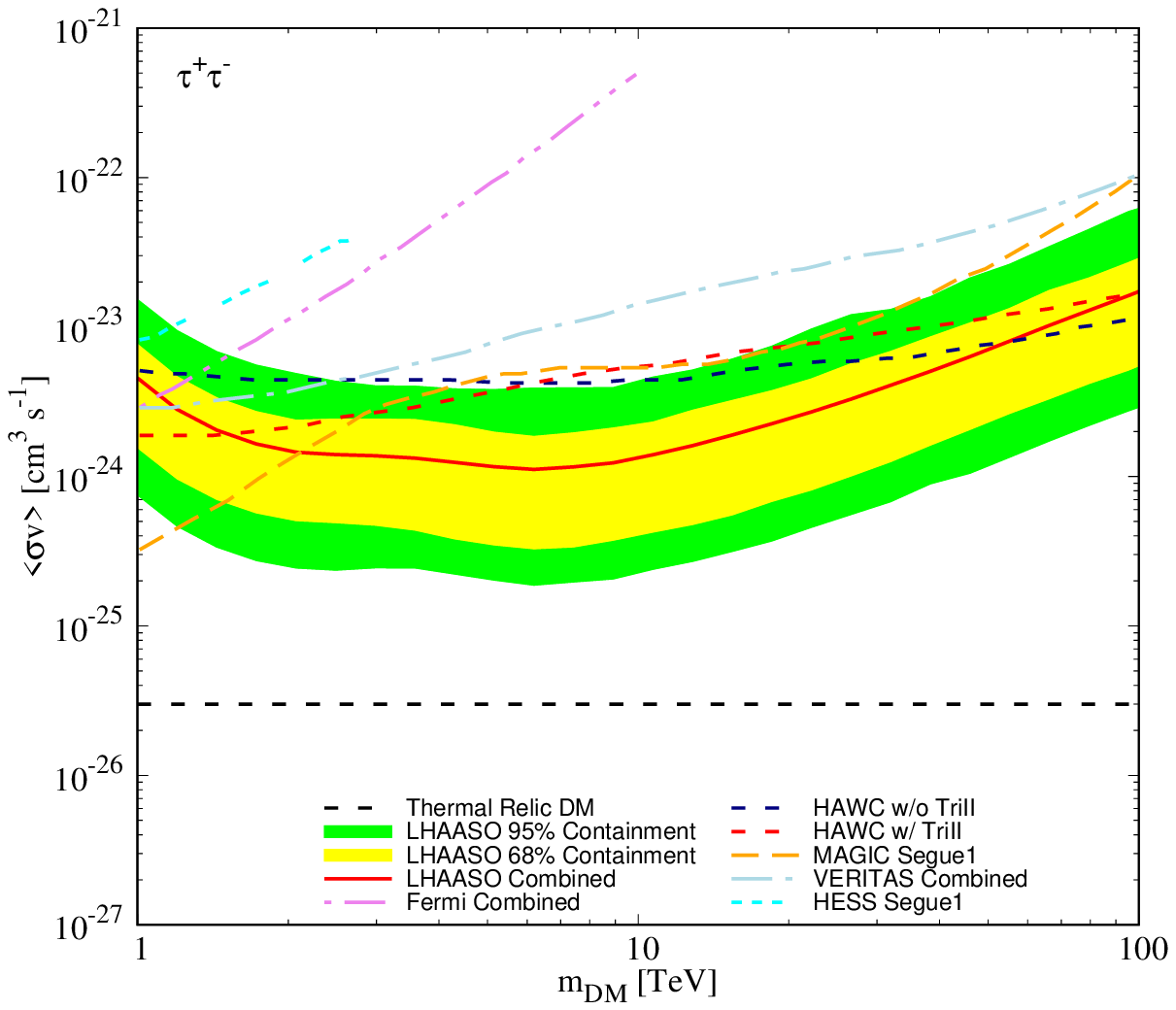}}
	{\includegraphics[width=0.45\textwidth]{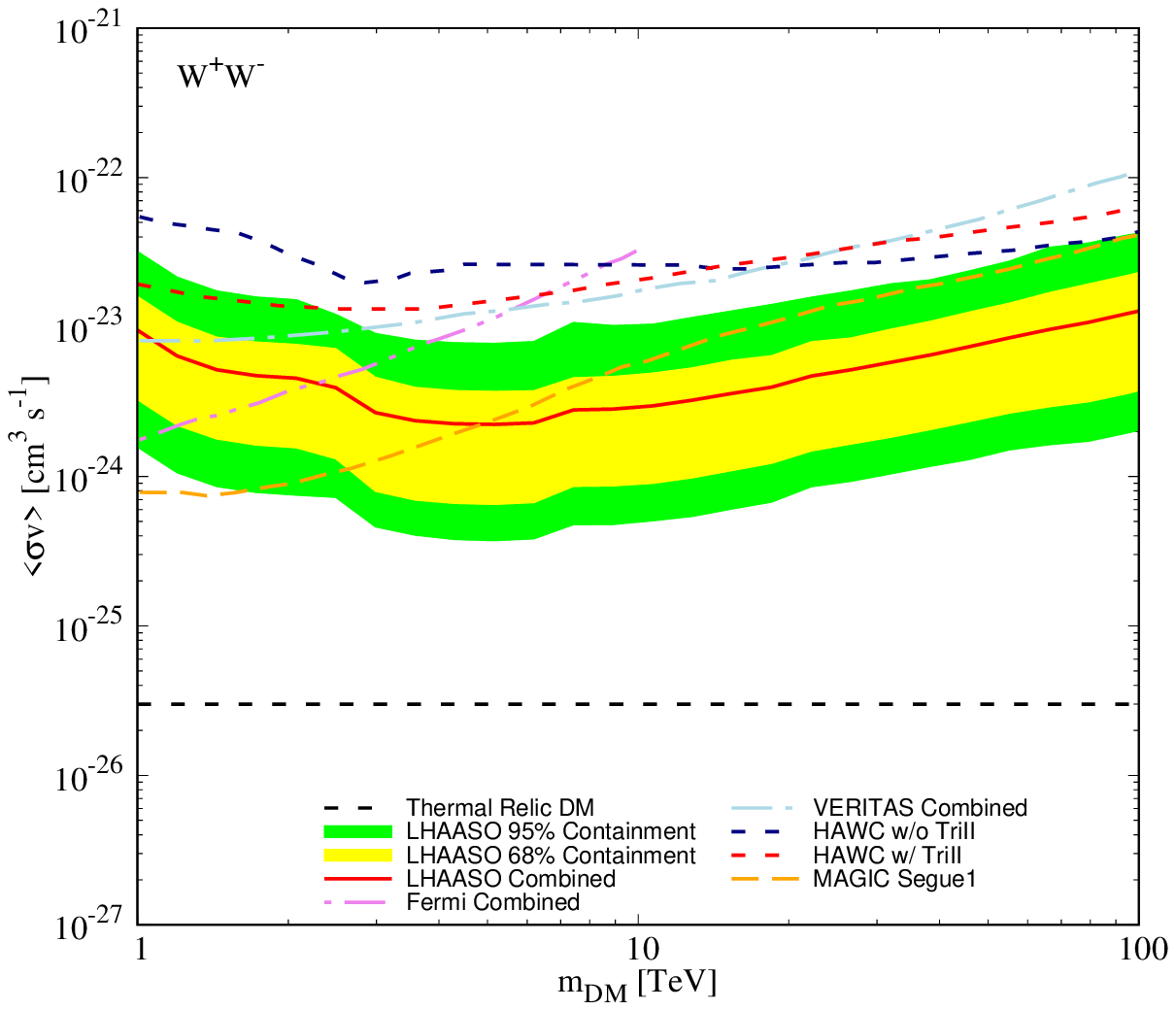}}
\caption{The LHAASO median combined sensitivities (red solid lines) and related two-sided $68\%$ (yellow bands) and $95\%$ (green bands) containment bands of one year for the b$\bar{\rm b}$, $\rm t\bar{\rm t}$, $\mu^{+}\mu^{-}$, $\tau^+\tau^-$, and $\rm W^+W^-$ annihilation channels. The HAWC combined limits \cite{Albert:2017vtb}, Fermi-LAT combined limit \cite{Ackermann:2013yva}, VERITAS Segue 1 limit \cite{Aliu:2012ga}, HESS combined dSph limit \cite{Abramowski:2014tra} and MAGIC Segue 1 limit \cite{Ahnen:2016qkx} are also shown for comparison. }
\label{fig:Lhaaso-combined}
\end{figure*}

To consider the statistic fluctuation in the analysis, we repeat 500 mimic observations under the null hypothesis considering the Poisson fluctuation on the expected event count. Then, we calculate the median combined sensitivity and the two-sided 68\% and 95\% containment bands as shown in Fig.~\ref{fig:Lhaaso-combined}.
In this figure, we also displayed the comparison of the LHAASO sensitivity to the constraints from another five dSph gamma-ray observations, including the HAWC combined limit \cite{Albert:2017vtb}, Fermi-LAT combined dSph limit \cite{Ackermann:2013yva}, HESS combined dSph limit \cite{Abramowski:2014tra}, VERITAS Segue 1 limit \cite{Aliu:2012ga} and MAGIC Segue 1 limit \cite{Ahnen:2016qkx}.

Broadly speaking, the most strong LHAASO sensitivity to the DM annihilation cross section comes from the $\tau^{+}\tau^{-}$ annihilation channel, which is nearly close to $10^{-24}~\rm cm^{3}~s^{-1}$, for all the DM masses considered here. For the $b\bar{b}$ channel, the MAGIC observation sets the most stringent constraint up to $\sim15$ TeV. Above $\sim15$ TeV, LHAASO is more sensitive to this channel. Besides, with regard to the $t\bar{t}$, $\tau^+\tau^-$, and $\rm W^+W^-$ channels, LHAASO are more sensitive beyond $\sim3 ~\rm TeV$, $\sim2 ~\rm TeV$, and $\sim 5 ~\rm TeV$, respectively, compared with the current limits set by other experiments.
For the $\mu^{+}\mu^{-}$ channel, LHAASO has great sensitivity for almost the whole mass range from 1 to 100 TeV. Therefore, we can conclude that LHAASO will be able to set stringent constraints on the property of heavy DM particles, especially for those heavier than $\sim 10$ TeV.

\section{Conclusion and discussion}\label{sec:conclu}
LHAASO is a newly planed under-construction wide FOV observatory to research the VHE gamma-ray astronomy with unprecedented sensitivity. Considering the fact that LHAASO will carry out its preliminary operation at the end of this year, it is timely to predict the physical perspective of LHAASO based on the simulated experimental data.
In this paper, we investigate the LHASSO sensitivity to the DM annihilation cross section for five DM annihilation channels by the gamma-ray observation of dSphs. We calculate the individual sensitivities for 19 dSphs within the LHAASO FOV using a likelihood ratio analysis method. To make the analysis more comprehensive and reliable, the statistical uncertainty of the $J$-factor is also incorporated as a nuisance parameter in the likelihood formulation.
In addition, we also calculate the combined sensitivity from a joint likelihood analysis of overall dSphs with the purpose of enhancing the statistical power in the calculation. These are the first simulated LHAASO sensitivities to the DM annihilation cross section using the mimic observation data.

Our calculation shows that the LHAASO combined sensitivity is dominated by the influence of the three dSphs with large $J$-factors: Segue 1, Ursa Major II, and Triangulum II. Furthermore, we compare the LHAASO sensitivities with the current limits set by other five gamma-ray experiments, including HAWC, Fermi-LAT, VERITAS, HESS and MAGIC.
The results manifest that the LHAASO sensitivities are better than the current limits above $\sim2$, 5, and 8 TeV for the $\tau^+\tau^-$, $\rm W^+W^-$ and $b\bar{b}$ channels, respectively. For the $t\bar{t}$ and $\mu^{+}\mu^{-}$ channels, LHAASO has great sensitivities in the large mass range from 1 to 100 TeV.

It is worthwhile to mention that several systematic uncertainties arising from the determination of the $J$-factor of dSphs would contribute a factor of several on the uncertainty of the final sensitivity.
In spite of the existence of these uncertainties, our results still show that the LHAASO gamma-ray research of dSphs would be a promising way for the DM indirect detection. It is believed that LHAASO will greatly enrich our knowledge about DM particles above ${\mathcal O}$(TeV).

\begin{acknowledgments}
We would like to thank the anonymous referees for their helpful comments and suggestions.
We are grateful to Ying-Ying Guo, Min Zha, Yan-Jin Wang, Yi-Qing Guo, Han-Rong Wu and Zhi-Guo Yao for helpful and engaging discussions.
This work is supported by the National Key R~$\&$~D Program of China (Grant No. 2016YFA0400200), the National Natural Science Foundation of China (Grants No. U1738209, No. 11851303, and No. 11835009), and the National Program for Support of Top-Notch Young Professionals.
\end{acknowledgments}

\appendix
\section{Gamma-ray absorption} \label{gamma_absorption}
The VHE gamma rays would be absorbed when traveling through the cosmos due to the $e^+e^-$ pair production.
This effect has been well studied in the literature \cite{Franceschini:2008tp,Gould:1967zzb}.
The differential optical depth $\mathrm{d}\tau/\mathrm{d}x$ of the cosmic photon gas depends on both the $\gamma+\gamma$ collision cross section and the number density of the target photons.

The $\gamma+\gamma\rightarrow e^++e^-$ cross section is
\begin{equation}
  \sigma = \frac{1}{2}\pi r_0^2(1-\beta^2)\left[ (3-\beta^4)\ln\frac{1+\beta}{1-\beta}-2\beta(2-\beta^2) \right],
  \label{eq:cross_section}
\end{equation}
where $r_0=e^2/mc^2$ is the classical electron radius and $\beta$ is the velocity of the $e^+$ ($e^-$)  in the center-of-mass system.

With an auxiliary variable $s\equiv (EE'/2m^2c^4)(1-\cos\theta)=1/(1-\beta^2)$ defined, the differential optical depth
\begin{equation}
  \frac{\mathrm{d}\tau}{\mathrm{d}x}=\int\int\frac{1}{2}\sigma n(E')(1-\cos\theta)\sin\theta\mathrm{d}E'\mathrm{d}\theta
  \label{eq:dtaudx}
\end{equation}
could be separated into two integrations
  \begin{equation}
    \left\{
      \begin{aligned}
        \frac{\mathrm{d}\tau}{\mathrm{d}x}&=\pi r_0^2\left( \frac{m^2c^4}{E} \right)^2\int^\infty_{m^2c^4/E}E'^{-2}n(E')\varphi\left[ s_0(E') \right]\mathrm{d}E'\\
        \varphi\left[ s_0(E') \right] &= \frac{2}{\pi r_0^2}\int_{1}^{s_0}\sigma(s)\mathrm{d}s
    \end{aligned}\right.,
  \label{eq:dtaudx_final}
  \end{equation}
  where $s_0\equiv 2s/(1-\cos\theta)$. This formula is more convenient for the analytical analysis compared to the original one. The reader is referred to Ref.~\cite{Gould:1967zzb} for more details.

For the CMB, the photon gas density follows a blackbody distribution
\begin{equation}
  n(E') = (\hbar c)^{-3}(E'/\pi)^2\left( e^{E'/kT}-1 \right)^{-1},
  \label{eq:CMB_density}
\end{equation}
where the temperature $T$ for CMB is approximately to be $2.73\,\mathrm{K}$.
Adopting the distribution in Eq.~(\ref{eq:CMB_density}), Ref.~\cite{Gould:1967zzb} has derived two asymptotic formulae for the solution of Eq.~(\ref{eq:dtaudx_final}).
Unfortunately, there was a typo in the Eq.(10) of Ref.~\cite{Gould:1967zzb}, where the term $-L(\omega_0)$ ought to be $-4L(\omega_0)$.
  As a result, the asymptotic formulae ought to be
  \begin{equation}
    \begin{aligned}
      f(\nu)&\rightarrow(\pi^2/3)\nu\ln(0.117/\nu), \nu\ll1 \\
      f(\nu)&\rightarrow\sqrt{\pi}e^{-\nu}(\sqrt{\nu}+6/\sqrt{\nu}\dots), \nu\gg1
    \end{aligned}.
    \label{eq:asymptotic}
  \end{equation}
  Both of the Eq.~(\ref{eq:asymptotic}) and the numerical integral could lead to reasonable results.

\begin{figure}[!htpb]
  \centering
  \includegraphics[width=0.7\textwidth]{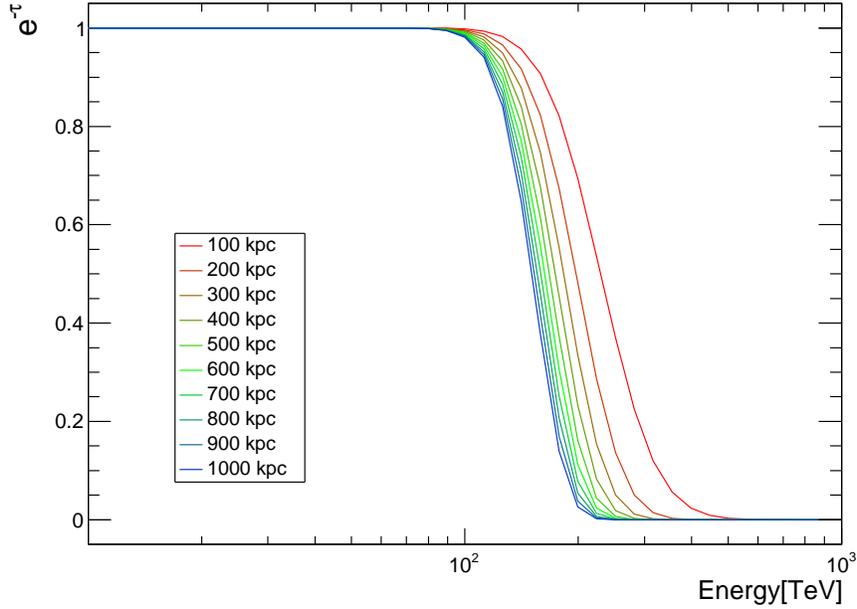}
  \caption{The survival rate of gamma-rays in CMB.}
  \label{fig:opacity}
\end{figure}

Here, we directly calculate the Eq.~(\ref{eq:dtaudx_final}) for CMB, and show the corresponding survival rate $e^{-\tau}$ for source distances $100\,\kpc\sim1\,\Mpc$ in Fig.~\ref{fig:opacity}.
The optical depth $\tau$ used here is just the product of the distance $d$ and the differential optical depth $\mathrm{d}\tau/\mathrm{d}x$.
It could be seen that even for a distance as far as $1\,\Mpc$, the survival rate of gamma-rays $\sim20\,\TeV$ is almost $100\%$.
Therefore, the absorption of CMB is negligible for our target sources whose distances are always smaller than $300\,\kpc$.

In addition to the CMB, the infrared radiation component of interstellar radiation field (ISRF) would also absorb the gamma-ray.
The infrared radiation is composed of multiple kinds of emission from the cosmic dust.
Its distribution inside the Galaxy is much more complex than that of the CMB.
A widely adopted detailed model has been established to describe this distribution in Ref.~\cite{Strong:1998fr}.

Reference~\cite{Zhang:2005tp} has adopted this model to perform a detailed calculation to analyse the gamma-ray attenuation for the $\mathrm{Sgr\,\,A^*}$ at Galactic center (GC), showing that the infrared radiation component would lead to an absorption $\lesssim5\%$ for the photon energies below $20\,\TeV$.
Note that the infrared radiation is supposed to be densest at the GC and would rapidly decrease along the axes of $R$ and $z$ in the cylindrical coordinate~\cite{Strong:1998fr}.
Therefore, the optical depth for the dSphs of interest in this work would be even much smaller than that of the $\mathrm{Sgr\,\,A^*}$.
We thus could neglect the absorption of infrared radiation inside the Galaxy.

\bibliographystyle{utphys}
\bibliography{reference}

\end{document}